\documentclass[11pt]{article}
\usepackage[utf8]{inputenc}
\usepackage[legalpaper,margin=1.0in]{geometry}
\usepackage{authblk}
\usepackage[round]{natbib}
\usepackage{graphicx}
\usepackage[caption=false,font=footnotesize]{subfig}
\usepackage{booktabs} 
\usepackage{cite}
\usepackage{apptools}
\usepackage{amsmath,amssymb,xcolor}
\usepackage{amsfonts}
\usepackage{amsthm,comment}
 \usepackage{algorithm} 
\usepackage{mathtools}
\usepackage{algcompatible}
\usepackage{enumitem}
\usepackage[noend]{algpseudocode}
\usepackage{url}
\usepackage{bm}
\def\D{\bm{D}}
\def\X{\bm{X}}
\def\Pb{\bm{P}}

\def\real{\mathbb{R}}
\def \alphab {\bm{\alpha}}
\def \betab {\bm{\beta}}
\def \v{\bm{v}}
\def \w{\bm{w}}
\def\S{\mathbb{S}}
\def \alphab {\bm{\alpha}}
\def \betab {\bm{\beta}}
\def \e{\bm{e}}
\def \f{\bm{f}}

\def\A{\bm{A}}
\def\B{\bm{B}}
\def\C{\bm{C}}
\def\D{\bm{D}}
\def\E{\bm{E}}
\def\F{\bm{F}}
\def\G{\bm{G}}
\def\H{\bm{H}}
\def\I{\bm{I}}

\def\K{\bm{K}}
\def\L{\bm{L}}

\def\R{\bm{R}}
\def\U{\bm{U}}
\def\V{\bm{V}}
\def\W{\bm{W}}
\def\X{\bm{X}}
\def\Y{\bm{Y}}
\def\Z{\bm{Z}}
\def\P{\bm{P}}
\def\d{\bm{d}}
\def\a{\bm{a}}
\def\b{\bm{b}}
\def\c{\bm{c}}
\def\d{\bm{d}}
\def\s{\bm{s}}
\def\x{\bm{x}}
\def\y{\bm{y}}
\def\Kn{\mathcal{K}_n}

\def\univ{\mathbb{I}}

\def\bg{\bigg}

\def\ones{\bm{1}}
\def\zeros{\bm{0}}
\def\snl{\vspace{\baselineskip}}

\newcommand*{\matminus}{%
  \leavevmode
  \hphantom{0}%
  \llap{%
    \settowidth{\dimen0 }{$0$}%
    \resizebox{1.1\dimen0 }{\height}{$-$}%
  }%
}

\renewcommand{\epsilon}{\varepsilon}

\newtheorem{theorem}{Theorem}
\newtheorem{corollary}{Corollary}

\DeclareMathOperator*{\minimize}{\mathrm{minimize}}
\DeclareMathOperator*{\subjectto}{\mathrm{subject~to}}

\begin{document}
\title{Localization from structured distance matrices via low-rank matrix recovery}
\date{\vspace{-4ex}}

\author[1]{Samuel Lichtenberg}
\author[1]{Abiy Tasissa}
\affil[1]{Department of Mathematics, Tufts University, Medford, MA 02155, USA
}

\maketitle

\begin{abstract}

We study the problem of determining the configuration of $n$ points by using their distances to $m$ nodes, referred to
as anchor nodes. One sampling scheme is Nystr\"om sampling, which assumes known distances between the anchors
and between the anchors and the $n$ points, while the distances among the $n$ points are unknown. For this scheme, a simple adaptation of the Nystr\"om method, which is often used for kernel approximation,
is a viable technique to estimate the configuration of the anchors and the $n$ points. In this manuscript, we propose
a modified version of Nystr\"om sampling, where the distances from every node to one central node are known, but all other distances are incomplete. In this setting, the standard Nystr\"om approach is not applicable, necessitating an alternative technique to estimate the configuration of the anchors and the $n$ points.
We show that this problem can be framed as the recovery of a low-rank submatrix of a Gram matrix.  
Using synthetic and real data, we demonstrate that the proposed approach can exactly recover configurations of points given sufficient distance
samples. This underscores that, in contrast to methods that rely on global sampling of distance matrices, the task of estimating the configuration of points can be done efficiently via structured sampling with well-chosen reliable anchors. Finally, our main analysis is grounded in a specific centering of the points. With this in mind, we extend previous work in Euclidean distance geometry by providing a general dual basis approach for points centered anywhere.

\end{abstract}

\section{Introduction}
\label{sec:intro}

Given incomplete pairwise distance information for a set of $n$ objects, a core challenge across diverse domains \citep{fang2013using,lavor2012recent,ding2010sensor,biswas2006semidefinite, einav2023quantitatively,porta2018distance,rojas2012distance}
is to recover the configuration of points realizing those distances, a problem known as the Euclidean distance geometry (EDG) problem. Besides partial distance information, we frequently have additional side information in practical applications.
In particular, we may already know the full or partial pairwise distances between a small subset of the points, or the missing distance values may exhibit a structured pattern that reflects the specific measurement protocol in use. In this paper, our focus is centered on the scenario when these assumptions hold. We illustrate this scenario through the localization problem.

Localization is a fundamental problem in sensor networks. Given a collection of sensor nodes deployed over a certain area,
the goal of localization is to obtain the positions of the sensors while employing protocols which minimize cost and power consumption \citep{kuriakose2014review}.
The location data is indispensable in a variety of applications, including but not limited to geographic routing \citep{hao2018integrating,li2012localized}, environmental monitoring \citep{verma2010wireless,liu2007wireless,hakala2008wireless}, and structural health monitoring \citep{balageas2010structural,bouzid2015structural,dos2014localized}. 
We refer the reader to \citep{mao2007wireless} for a survey of localization algorithms. 
We discuss one setting of localization in which we are given the full pairwise distance matrix between $m$ nodes. For the remaining $n$ sensor nodes,  we are provided information in terms of pairwise distances to the $m$ nodes. We adopt the established terminology in localization, henceforth referring to these nodes as ``anchor nodes'' and ``mobile nodes'' respectively. 

Formally, let $\x_1,...\x_m$ and $\y_{1},...,\y_{n}$ denote
the positions of $m$ anchor nodes and $n$ mobile nodes respectively. Here on, we assume the points lie in $\real^r$, with $r$ typically set to $2$ or $3$ in many applications. The underlying squared pairwise distance matrix for the $p = m+n$ sensor nodes has the following form
\begin{align}
\label{eq:dist_block_structure}
  \D = \left[
  \arraycolsep=4pt\def\arraystretch{1.6}
  \begin{array}{ c | c }
    \E & \F \\
    \hline
    \F^{\top} & \G
  \end{array}\right],
\end{align}
where $\E \in \real^{m\times m}$ denotes the anchor-anchor squared distance matrix, $\F \in \real^{m\times n}$
denotes the anchor-mobile squared distance matrix and $\G \in \real^{n\times n}$ is the mobile-mobile squared distance
matrix. One sampling scheme is the scheme used in the Nystr\"om method \citep{NIPS2000_19de10ad,platt2004nystrom}, where we have access to $\E$ and $\F$ but lack distance information between the mobile nodes, meaning no entries of $\G$ are observed. We refer to this sampling model as Nystr\"om sampling. The problem is then to estimate the configuration of
the $p$ nodes using the distances in the blocks $\E$ and $\F$. This setup differs slightly from the standard Nystr\"om problem in that $\D$ is not a kernel matrix, and not the main object we are trying to recover. However, following the analysis in \citep{platt2004nystrom}, this problem can be translated into a standard Nystr\"om problem on the Gram (kernel) matrix $\K$. Specifically, if we set $\P= [\X \,\, \Y]$ with $\X=
[\x_1,\x_2,...,\x_m]$ and $\Y = [\y_1,...,\y_n]$, the Gram matrix associated to $\P$ is defined as $\K = \P^T\P$ and has the following block form
\begin{align}
\label{eq:gram_block_structure}
   \K = \left[
      \arraycolsep=4pt\def\arraystretch{1.6}
    \begin{array}{ c | c }
    \X^T\X & \X^T\Y \\
    \hline
    \Y^{T}\X & \Y^T\Y 
  \end{array}\right]
     = \left[
      \arraycolsep=4pt\def\arraystretch{1.6}
    \begin{array}{ c | c }
    \A & \B \\
    \hline
    \B^{\top} & \C 
  \end{array}\right].
\end{align}
Here, $\A\in \real^{m\times m}$ and $\B\in \real^{m\times n}$ are considered fully known, whereas $\C\in \real^{n\times n}$ is unknown. Specifically, $\B$ and $\A$ can be derived from $\E$ and $\F$ via relationships given in \citep{platt2004nystrom}, and then $\C$ can then be recovered via the usual Nystr\"om method on kernel matrices \citep{NIPS2000_19de10ad}. The Nystr\"om sampling is very useful for problems with structured partial distance information, but it requires that all distance entries of $\E$ and $\F$ are known in order to recover $\B$. In practice, this may be restrictive or expensive as it requires communication between a given anchor node and all mobile nodes. 

In this paper, we propose a modified Nystr\"om sampling scheme that addresses missing distances in the $\E$ and $\F$ blocks. This approach is termed anchorless, as it does not require the positions of the anchors and could be more cost-effective than anchor-based methods \citep{popescu2012manifold, kulmer2018anchorless}.
The main motivation for our sampling scheme is to offer flexibility to capture the various uncertainties in distance measurements across applications. In this context, anchors represent the subset of points for which we have reliable 
pairwise distance measurements, while mobile nodes signify points where the distance measurements are more likely to be uncertain. For example, in structure prediction, anchors could correspond to a  subset of atoms in a protein region that is easier to predict.

We now discuss theoretical analysis and algorithms related to this problem. Assuming that the entries of $\E$ and $\F$ 
are sampled from the uniform or Bernoulli sampling model, one approach is to use standard matrix completion algorithms \citep{candes2009exact, gross2011recovering, recht2010guaranteed, jain2013low, wei2020completion} to complete $\E$ and $\F$. However, these methods do not consider that $\E$ is a squared Euclidean distance matrix and $\F$ is a submatrix of such a matrix, thus ignoring non-negativity and triangle inequality constraints. The studies in \citep{lai2017solving, abiy_exact,li2024sensor} note that using a Gram matrix formulation offers optimal sampling complexity (number of distance entries needed for exact recovery
of the points) for the EDG problem, in contrast to a distance matrix formulation, which does not. This is due to the Gram matrix's structure, which is symmetric and positive semi-definite, and an essential equivalence between Gram matrices and distance matrices \citep{schoenberg1935remarks}. Therefore, an approach would be to use the method in \citep{abiy_exact}, which deals with completing a distance matrix from randomly sampled entries. However, the random sampling in \citep{abiy_exact} is global, unlike our setting, which only samples in the $\E$ and $\F$ blocks. Thus, the theoretical analysis in \citep{abiy_exact} is not directly applicable here. Additionally, existing works \citep{li2024sensor, nguyen2016matrix, smith2023riemannian} that consider random sampling of the distance matrix without assuming any special structure cannot be applied to our setup. 

The closest work to ours is \citep{cai2023matrix}, which considers the problem of completing a low-rank matrix with observations restricted to a few rows and columns. However, applying this method to a distance matrix requires integrating the properties of the distance matrix into the optimization, which potentially could require more distance samples. 
One of the main motivations of this paper is to demonstrate that, under the modified Nystr\"om sampling scheme, the localization problem can be formulated as a generalized matrix completion problem of the matrix $[\A\,\,\B]$, which is a submatrix of $\K$. Given that positive semidefinite matrices arise in many applications, the problem of completing such matrices from a few of their entries has been explored both algorithmically and theoretically \citep{alfakih1999solving, graepel2002kernel, laurent2014positive, mavroforakis2017active, NIPS2014_56468d56, scholkopf1997kernel}. While these works and our approach share the common idea of using the Gram matrix, these works focus on completion problem from sampled entries of the Gram matrix, whereas in the problem we consider, the measurements are based on the entries of the distance matrix.

\subsection{Contributions}

The key contributions of this paper are:
\begin{enumerate}

\item \textbf{Modified Nystr\"om sampling model}: We propose a sampling model where distance measurements are available between all mobile nodes and a single anchor node, meaning we have a complete row of the matrix $\F$. We refer to this anchor as the central node. For the other entries in $\E$ and $\F$, we assume partial pairwise distance information. Figure \ref{fig:sampling_illustration} shows an illustration of the sampling scheme. 
In our numerical experiments, the entries of $\E$ are sampled according to the Bernoulli model. For each column of $\F$, corresponding to a mobile node, we sample uniformly at random $\alpha$ entries. The interpretation of this is that we know the distance of each mobile node to $\alpha$ randomly selected anchor nodes. Our numerical experiments demonstrate that, provided well-chosen anchors and sufficient distance samples in $\E$ and $\F$ block, we can estimate the positions of the anchors and mobile nodes  efficiently. 
\begin{figure}[t!]
    \centering    \includegraphics{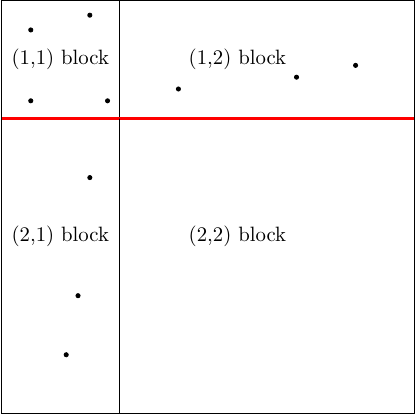}
    \caption{Illustration of the proposed sampling model. We sample few distances from the anchor-anchor and anchor-mobile squared distance matrix. The distances between the mobile nodes are not sampled. In addition, we assume that we know the distance of all mobile nodes from one of the anchors.
    Grey indicates the distances are known in that block while white denotes the distance information is missing in that block. The red line indicates the central node from which we know the distances to all mobile nodes. The black dots indicate observed distance entries in these blocks. }
    \label{fig:sampling_illustration}
\end{figure}

\item \textbf{Anchorless localization}: Under the modified Nystr\"om sampling model, we formulate the anchorless localization problem as the recovery of a low-rank matrix, specifically the submatrix $[\A\,\B]$ of $\K$, given a subset of its expansion coefficients in a non-orthogonal basis. The expansion coefficients depend on the observed entries of $\E$ and $\F$. A consequence of our derivation is that the central node in 
the modified Nystr\"om sampling model is necessary to relate the observed entries of $[\E\,,\F]$ to $[\A\,,\B]$.

\item \textbf{Centering analysis}: The analysis in this paper leverages a special centering of the points. Motivated by that, we extend previous analysis 
in \citep{lichtenberg2023dual} for Euclidean distance geometry by considering arbitrary centerings of the set of points.

\end{enumerate}

\section{Notation}

All vectors and matrices are represented in bold font. Given a vector $\x$, its $i$-th entry is denoted by
$x_i$. For a matrix $\X$, the $(i,j)$-th entry is denoted by $X_{i,j}$. The $i$-th row and $j$-th column 
of the matrix $\X$ are denoted by $\X(i,:)$ and $\X(:,j)$ respectively. $\X_{\mathcal{I},:}$ denotes the submatrix of $\X$ with row indices  $\mathcal{I}$. In a similar manner, $\X_{:,\mathcal{J}}$ denotes the submatrix of $\X$ with column indices $\mathcal{J}$. The identity matrix is denoted as $\I$. Standard basis vectors and matrices are represented by $\e_i$ and $\e_{\alphab}$ or $\e_{i,j}$. $\univ$ is the universal set $\{ (i,j): 1 \leq i < j \leq n \}$ of indices. $\ones$ is the all-ones vector and $\zeros$ is the all-zeros vector. 
The trace inner product for matrices is expressed as $\langle \X, \Y \rangle$. $\A \succeq \bm{0}$ denotes that $\A$ is a symmetric and positive semidefinite matrix. The delta function $\delta_i^j$ is defined as 1 when $i = j$ and 0 otherwise. $\lVert \cdot \rVert_{*}$ denotes the nuclear norm of a matrix. $\A^{\dagger}$ denotes the pseudo-inverse of $\A$. Occasionally, we may refer to a matrix block as $\v_{\alphab}^{(1,2)}$, representing the $(1,2)$ block of $\v_{\alphab}$. In the context of this work, focused on Nystr\"om methods, matrices exclusively consist of four blocks: (1,1), (1,2), (2,1), and (2,2). If these blocks pertain to the kernel matrix $\K$ or the squared distance matrix $\D$, they will be assigned special names as will be detailed in the Background section.

\section{Background}

This section offers a concise overview of the Nystr\"om method and the dual basis approach, which serve as foundational concepts for the subsequent sections.
\subsection{The Nystr\"om Method for Gram Matrices}

The Nystr\"om method considers a low-rank approximation of a symmetric and positive semidefinite matrix $\K$, which has the block structure shown in \eqref{eq:gram_block_structure}.
It is typically assumed that $m \ll n$, and that we do not know entries of $\C$. If $\text{rank}(\A) = \text{rank}(\K)$, then every column $\left[ \begin{array}{ c } \B(:,j) \\ \C(:,j) \end{array}\right]$, $1 \leq j \leq n$, can be written as a linear combination of columns
$\left[ \begin{array}{ c } \A(:,i) \\ \B^{\top}(:,i) \end{array}\right], 1 \leq i \leq m$. 
In other words, $\B(:,j) = \A \x_j$ and $\C(:,j) = \B^{\top} \x_j$, for some $\x_j \in \real^{m}$.
This gives the equations $\B = \A \X$ and $\C = \B^{\top} \X$ for some $\X \in \real^{m \times n}$. If $\A$ is invertible, then we must have $\C = \B^{\top}\A^{-1}\B$. 
If $\A$ is not invertible, then $\X = \A^{\dagger}\B$ provides the least-squares solution to the first equation, and so the Nystr\"om method sets $\C = \B^{\top}\A^{\dagger}\B$. This agrees with the prior equation for $\C$ when $\A$ is invertible. If $\K$ is at most rank $m$ and $\textrm{rank}(\A) = \textrm{rank}(\K)$, then the Nystr\"om approximation is exact \citep{kumar2009sampling}.
An essential problem is determining the block structure of $\K$ to ensure the Nystr\"om method is exact. In \citep{talwalkar2014matrix}, it is noted that if $\A$ is formed by uniformly sampling the rows (and their corresponding columns) of $\K$ with replacement, the number of rows in $\A$, denoted as $m$, depends on a specific quantity known as coherence of $\K$ (see Definition $1$ in \citep{talwalkar2014matrix}).

\subsection{The Nystr\"om Method for Distance Matrices}
Suppose we have a squared distance matrix $\D$ and a kernel (Gram) matrix $\K$ with the structures shown in \eqref{eq:dist_block_structure} and \eqref{eq:gram_block_structure}.
We presume that the partition into sub-blocks occurs within the initial $m$ rows and columns. The objective is to recover a $\K$ that generates $\D$ by observing only a portion of $\D$, specifically all the entries in $\E$ and $\F$. This is achieved by exploiting the connection between $\D$ and $\K$ as outlined below. For any vector $\s$ that sums to $1$, the works in \citep{gower1982euclidean} and \citep{gower1985properties} demonstrate the following relation between $\K$ and $\D$:
\begin{equation}\label{eq:k_d_relationship}
\K = -\frac{1}{2}(\I - \ones \s^{\top})\D(\I - \s \ones^{\top}).
\end{equation}
This procedure, when applied to $\D$, is often called double-centering. Different choices of $\s$ will yield different $\K$, but any $\K$ obtained in this way will always reproduce $\D$. 

The equation in \eqref{eq:k_d_relationship} may be expanded for the $(i,j)$-th entry of $\K$ as follows:
\begin{align*}
    K_{i,j} = -\frac{1}{2}\bg(D_{i,j} - \sum_{q=1}^N s_q D_{i,q} &- \sum_{p=1}^N s_p D_{p,j} + \sum_{p=1}^N \sum_{q=1}^N s_p s_q D_{p,q}\bg).
\end{align*}
By choosing $\s$ to be $1/m$ in the initial $m$ entries and $0$ elsewhere, as described in \citep{platt2004nystrom}, we get the following result:
\begin{equation}
\begin{aligned}
\label{eq:platt_a}
    A_{i,j} = -\frac{1}{2}\bg( E_{i,j} - \frac{1}{m}\sum_{p=1}^m E_{p,j} &- \frac{1}{m}\sum_{q=1}^m E_{i,q} + \frac{1}{m^2}\sum_{p=1}^m \sum_{q=1}^m E_{p,q} \bg) 
\end{aligned}
\end{equation}
and
\begin{equation}
\begin{aligned}
\label{eq:platt_b}
     B_{i,j} = -\frac{1}{2}\bg( F_{i,j} - \frac{1}{m}\sum_{p=1}^m F_{p,j} &- \frac{1}{m}\sum_{q=1}^m E_{i,q} + \frac{1}{m^2}\sum_{p=1}^m \sum_{q=1}^m E_{p,q} \bg).   
\end{aligned}
\end{equation}

By choosing a particular centering vector $\s$, we are able to recover the $\A$ and $\B$ blocks of $\K$ using only the $\E$ and $\F$ blocks of $\D$. A standard Nystr\"om procedure for Gram matrices can then be performed to recover all of $\K$. From the solution of $\K$, akin to classical multidimensional scaling 
\citep{young1938discussion,torgerson1952multidimensional,torgerson1958theory,gower1966some}, we can recover the configuration of points using eigendecomposition.

\subsection{The Dual Basis Approach for Low-Rank Matrix Recovery}
\label{sec:dual_basis_approach}
The matrix completion problem aims to reconstruct a low-rank matrix $\X$ based on sampled measurements $\langle \X, \w_{\alphab} \rangle$, where $\{\w_{\alphab}\}$ is a set of orthogonal measurement operators \citep{recht2010guaranteed, gross2011recovering}. Orthogonality allows any $\X$ in $\text{span}(\{\w_{\alphab}\})$ to be represented as:
\begin{align}
\label{eq:expansion_of_X}
\X = \sum_{\alpha \in \univ} \langle \X, \w_{\alphab} \rangle \w_{\alphab}.
\end{align}
In this representation, ${\langle \X, \w_{\alphab} \rangle}$ are the expansion coefficients. The matrix completion problem can equivalently be formulated as the problem of determining a low-rank matrix $\X$ given a subset of its expansion coefficients. For instance, in the standard matrix completion problem where only a few entries of $\X$ are observed, $\{\w_{\alphab}\}$ would be the canonical basis, and $\{\langle \X, \w_{\alphab} \rangle\}$ would be the set of observed entries.

In the Euclidean distance geometry problem, each observation $D_{i,j}$ is related to the Gram matrix $\K$ as follows: $D_{i,j} = K_{i,i} + K_{j,j} - K_{i,j}-K_{j,i}$. For this case, the problem of completing the Gram matrix can not be represented as in \eqref{eq:expansion_of_X} where the expansion coefficients are the measurements. The main reason is that the inherent measurement operators are not orthogonal. Alternatively, a biorthogonal dual basis can be employed, as developed in \citep{abiy_exact}. A biorthogonal dual basis is a pair of indexed sets of linearly independent vectors $\{ \w_{\alphab} \}, \{ \v_{\betab} \}$, satisfying $\langle \v_{\alphab}\,,\w_{\betab} \rangle = \delta_{\alpha,\,\beta}$ for every $\alpha, \beta.$ 
We also sometimes use the phrase ``dual basis" to refer specifically to the set $\{ \v_{\alphab} \}$, which, in finite dimension, is uniquely determined by the basis set $\{ \w_{\alphab} \}$. 
When $\{ \w_{\alphab} \}$ is orthogonal, then $\v_{\alphab} = \w_{\alphab}$, so a biorthogonal dual basis is a generalization of an orthogonal basis. 
Biorthogonality ultimately allows one to obtain a representation of any Gram matrix $\K \in \text{span}(\{ \w_{\alphab} \})$ as:
\begin{align*}
    \K = \sum_{\alphab \in \univ} \langle \K, \w_{\alphab} \rangle \v_{\alphab}.
\end{align*}
The paper \citep{abiy_exact} gives concrete details of how this representation is achieved and how it can be used to provide recovery guarantees for the EDG problem. Here, we will only give a brief overview of the approach. 

The dual basis approach formulates the EDG problem as a matrix recovery task using the operator basis: 
\begin{align*}
    \w_{\alphab} = \e_{\alpha_1,\alpha_1}+\e_{\alpha_2,\alpha_2}-\e_{\alpha_1,\alpha_2}-\e_{\alpha_2,\alpha_1}, 
\end{align*}
for $\alphab = (\alpha_1, \alpha_2), \alpha_1 < \alpha_2$.
We observe that there are $L = n(n-1)/2$ basis elements, the set $\{ \w_{\alphab} \}$ 
spans a linear subspace 
$\S = \{\Z \in \real^{n \times n}  \vert \Z = 
\Z^T,~\Z \cdot \ones = \zeros\}$ of dimension $L$, and   
the measurements $D_{i,j}$ correspond to the inner products $\langle \X, \w_{i,j} \rangle$. However, since $\{ \w_{\alphab} \}$ is not orthogonal, standard recovery results from the matrix completion cannot be used to prove recovery guarantees for optimization procedures with respect to this basis. However, \citep{abiy_exact} proved guarantees when a biorthogonal dual basis is used instead, and provided the form of the dual basis corresponding to the operator basis $\w_{\alpha}$.  
Specifically, 
given $\{ \w_{\alphab} \}_{\alpha = 1}^L$, define the matrix $\H$ as  $\H_{\alpha,\,\beta} = \langle \w_{\alphab}\,,\w_{\betab}\rangle$; the set of matrices $\v_{\alphab}= \sum_{\beta} \H^{-1}_{\alpha,\,\beta} \w_{\betab}$ then forms a dual basis to $\{ \w_{\alphab} \}$ satisfying $\langle \v_{\alphab}\,,\w_{\betab} \rangle = \delta_{\alpha,\,\beta}$, which is the definition of biorthogonality. Figure \ref{fig:basis-objects} gives concrete examples of these objects. We note that the general problem of completing a low-rank matrix given a few of its expansion coefficients in a dual basis has been studied in \citep{tasissa2021low}.

\begin{figure}[h!]
    \centering
    \begin{align*}
    &\w_{1,2} = \begin{bmatrix*}[r]
    1 & \matminus 1 & 0 & 0 \\ 
    \matminus 1 & 1 & 0 & 0 \\
    0 & 0 & 0 & 0 \\
    0 & 0 & 0 & 0
    \end{bmatrix*}
    \hspace{24pt}
    && \v_{1,2} = \frac{1}{16}\begin{bmatrix*}[r]
        3 & \matminus5 & 1 & 1 \\
        \matminus5 & 3 & 1 & 1 \\
        1 & 1 & \matminus 1 & \matminus 1 \\
        1 & 1 & \matminus 1 & \matminus 1
    \end{bmatrix*}
    \end{align*}
    \caption{Dual basis objects $\w_{1,2}, \v_{1,2}$ for $n=4$.}
    \label{fig:basis-objects}
\end{figure}
In the dual basis expansion, $\K = \sum_{\alpha} \langle \K\,,\w_{\alphab}\rangle\v_{\alphab}$, transforming  the EDG problem into the recovery of a low-rank matrix $\X$ of rank $r$, given a few of its expansion coefficients. A computational challenge associated with the dual basis approach is determining an explicit form for $\v_{\alphab}$, since the direct approach depends on inverting a matrix of size $O(n^2)$. 
For the EDG problem, assuming that the points are centered at the origin, the explicit form of $\v_{\alphab}$ is given in  \citep{lichtenberg2023dual}. This manuscript generalizes the analysis in \citep{lichtenberg2023dual} and provides explicit forms of the basis $\{\w_{\alphab}\}$ and $\{\v_{\alphab}\}$ by considering arbitrary centerings of the points. 

\section{Related Work}

\subsection{CUR Decomposition}

In CUR decomposition, the objective is to obtain a low-rank approximation of a matrix by utilizing selected rows and columns of the matrix \citep{hamm2020perspectives,mahoney2009cur}. Formally, let $\X$ represent the underlying matrix, with sampled row and column indices denoted as $\mathcal{I}$ and $\mathcal{J}$, respectively. The matrix $\C$ is constructed from the columns of $\X$ and is defined as $\C = \X_{:,\mathcal{J}}$. Similarly, $\R$ is formed from the rows of $\X$ and is defined as $\R = \X_{\mathcal{I},:}$. The CUR approximation of $\X$ is expressed as $\tilde{\X} = \C \U \R$, where $\U$ is the matrix that is intersection of $\R$ and $\C$. The quality of this approximation, indicating the difference between $\X$ and $\tilde{\X}$ in a suitable norm, relies on the chosen sampling scheme for selecting rows and columns.

Various sampling methods and their associated approximation guarantees have been explored  \citep{chiu2013sublinear,cur_adaptive_sampling,drineas2008relative,hamm2020stability,monte_carlo_cur,sorensen2016deim,voronin2017efficient}. Addressing the limitation of conventional CUR algorithms that assume a fully observed matrix, the authors in \citep{xu2015cur} introduce a CUR algorithm designed for partially observed matrices. However, this method still requires observing all entries in the sampled rows and columns. In \citep{cai2023matrix}, the authors propose a novel sampling model that interpolates between the uniform sampling model and CUR sampling. Notably, this sampling model offers flexibility by not necessitating the observation of all entries in the selected rows and columns. In the context of the problem we study in this paper, we could apply this algorithm to a partial distance matrix (which is illustrated in Figure \ref{fig:sampling_illustration}). It is important to highlight that the partial squared distance matrix has a rank of at most $r+2$ (compared to the rank of $\B$ which is at most $r$). This could imply increased sampling and potentially less favorable sampling complexity when working in the space of distance matrices (see \citep{lai2017solving, abiy_exact,li2024sensor}). Additionally, it is crucial to note that the sampling method in \citep{cai2023matrix} cannot be directly applied to the Gram matrix $\K$ due to the non-preservation of the sampling structure when mapping from a distance matrix to a Gram matrix. 

Finally, it is worth mentioning that the Nystr\"om method is a more specific instance of the CUR method, constrained to a symmetric and positive semidefinite target matrix. For different sampling schemes for the Nystr\"om method, we refer the reader to \citep{gittens2013revisiting,kumar2012sampling}.
In contrast to the standard Nystr\"om method, our sampling model is more flexible, as we do not assume full knowledge of all entries of $\A$ and $\B$, the $(1,1)$ and $(1,2)$ blocks of our Gram matrix.

\subsection{Matrix Completion}
Matrix completion is the problem of reconstructing a low-rank matrix based on a subset of its entries \citep{candes2010matrix,candes2009exact,candes2010power,keshavan2010matrix,keshavan2010matrix2}. Theoretical analyses in \citep{recht2011simpler,chen2015incoherence} employ a convex optimization program and demonstrate that, under mild assumptions, a $p \times p$ matrix of rank $r$ can be successfully recovered from $O(pr\log^2(p))$ randomly sampled entries with high probability. It is important to note that the aforementioned studies assume a uniform sampling model or Bernoulli sampling.
Other works, such as \citep{gross2011recovering,recht2010guaranteed}, extend the scope to consider observations not as individual entries but as inner products of the underlying matrix with pre-specified matrices. Consequently, one natural approach for solving the main problem in this paper is to apply the standard matrix completion algorithms to the submatrix $[\E\,\,\,\F]$. 
However, these methods do not leverage the fact that $[\E\,\,\,\F]$ is a submatrix of an underlying squared Euclidean distance matrix.
It is worth noting that there are related works addressing similar themes to ours, such as those considering non-uniform sampling \citep{wan2018matrix}, matrix completion with side information \citep{ledentorthogonal}, and matrix completion where the cost of obtaining observations is also taken into account \citep{wan2018matrix}.

\subsection{Euclidean Distance Geometry}
The Euclidean Distance Geometry problem (EDG) is concerned with recovering the positions of $n$ points in $\real^r$ based on a few observed entries from an $n\times n$ Euclidean distance matrix \citep{dokmanic2015euclidean,liberti2014euclidean}. In the setting of the localization problem, which is the focus of this paper, various approaches based in EDG have been explored \citep{biswas2006semidefinite,ding2010sensor,fliege2019euclidean,sremac2019noisy}. This paper aims to demonstrate that the localization problem, considering the proposed sampling model (refer to Figure \ref{fig:sampling_illustration}), is a generalized matrix completion problem. This realization leads to the development of an algorithm formulated solely in terms of the submatrix $[\A\,\B]$. A similar methodology, focusing on a matrix completion approach for recovering the underlying Gram matrix, is presented in \citep{abiy_exact}. Building upon this, the work in \citep{kojoian2023distance} investigates a similar sampling model and proposes a convex program for recovering the mobile nodes. It is noteworthy that, in contrast to \citep{kojoian2023distance}, where the model assumes a complete $\E$ and the optimization involves the full Gram matrix $\K$, our formulation handles incomplete $\E$ and the optimization in this paper relies solely on $\A$ and $\B$. This approach can potentially facilitate computation for large-scale problems.

\section{Dual Basis Formulation}

\subsection{Relating Columns of $\F$ and $\B$ via the Graph Laplacian}

Before beginning this section, we introduce some brief preliminaries on graph Laplacians. For an unweighted graph $G = (V, E)$ on $n$ vertices, where $V$ denotes the set of vertices and $E$ denotes the set of edges, the graph Laplacian $L_G \in \real^{n \times n}$ is a symmetric matrix where $L_G(u,v) = -1$ if $(u,v) \in E$ (0 otherwise), and $L_G(u,u) = \text{deg}(u)$, the degree of $u$ i.e., the number of edges incident to $u$ \citep{chung1997spectral}. Of particular relevance to our discussion is the scenario when $G$ is the complete graph $\Kn$. In this case, every diagonal entry of $L_G$ is $n-1$, and every off-diagonal entry is $-1$.

\begin{theorem}
\label{b_laplacian_f_claim}
 Let $\L$ be the Laplacian of the complete graph $\mathcal{K}_m$, and $\b_j$ be a column of $\B$. Define the vector $\tilde{\f}_j$ as follows: $(\tilde{\f}_j)_s  = -\frac{1}{2m}(F_{s,j} - \frac{1}{m}\sum_{t=1}^m E_{s,t})$, for $s \leq m$. Then
\begin{align*}
    \b_j = \L \tilde{\f}_j.
\end{align*}
\end{theorem}

\begin{proof}

Using \eqref{eq:platt_b}, we express the $j$-th column of $\B$ as:
\begin{equation*}
    \begin{split}
    \b_j &= -\frac{1}{2}\bigg[\sum_{s=1}^{m} F_{s,j}\e_s - \left(\frac{1}{m}\sum_{s=1}^{m} F_{s,j}\right)\ones 
    - \sum_{s=1}^{m}\left( \frac{1}{m}\sum_{t=1}^{m} E_{s,t}\right)\e_s
    +\frac{1}{m^2}\left(\sum_{s=1}^{m}\sum_{t=1}^{m} E_{s,t}\right)\ones\bigg].
    \end{split}
\end{equation*}

We can further simplify the above expression as follows:
\begin{align*}
    \b_j &= -\frac{1}{2}\bg(\sum_{s=1}^{m} F_{s,j}\e_s - \bg(\frac{1}{m}\sum_{s=1}^{m} F_{s,j}\bg)\ones -\sum_{s=1}^{m}\bg(\frac{1}{m}\sum_{t=1}^{m} E_{s,t}\bg)\e_s
    +\frac{1}{m^2}\bg(\sum_{s=1}^{m}\sum_{t=1}^{m} E_{s,t}\bg) \ones \bg)\\
    &= -\frac{1}{2}\bg(\sum_{s = 1}^m F_{s,j}(\e_s - \frac{1}{m}\ones) -\frac{1}{m} \sum_{s=1}^m \sum_{t=1}^m E_{s,t}(\e_s - \frac{1}{m}\ones) \bg) \\
    &= -\frac{1}{2}\sum_{s=1}^m \bg((F_{s,j} - \frac{1}{m}\sum_{t=1}^m E_{s,t})(\e_s - \frac{1}{m}\ones)\bg) \\
    &= \sum_{s=1}^m \bg( -\frac{1}{2}\big(F_{s,j} - \frac{1}{m}\sum_{t=1}^m E_{s,t}\big)\big(\frac{1}{m}\L(:,s)\big)\bg) \\
     &= \sum_{s=1}^m [\tilde{\f}_j]_s\L(:,s) \\   
    &= \L \tilde{\f}_j.
\end{align*}
\end{proof}
Utilizing the above claim, we quickly derive two valuable observations. It is worth noting that the Laplacian relationship is not essential for demonstrating these corollaries; both could have been inferred directly from \eqref{eq:platt_b} alone, albeit with slightly more intricate equations.

\begin{corollary}
\label{b_col_sum_zero}
Every column $\b_j$ of $\B$ sums to zero.
\end{corollary}
\begin{proof}
Using the above claim, we observe that
 \begin{align*}
    &\b_j = \L \tilde{\f}_j \implies
    \b_j^{\top} \ones = \tilde{\f}_j^{\top} \L \ones = \zeros,
\end{align*}
where the last equation follows from the fact that $\L \ones = \zeros$. 
\end{proof}
In \eqref{eq:platt_b}, it is evident that a straightforward correspondence between an entry of $\B$ and an entry of $\F$ does not exist. Our second observation, however, reveals a clear correspondence when differences of entries are considered instead.

\begin{theorem}
\label{b_f_entries_cor}
For $i,k \in [1,m]$, we have
\begin{align*}
    B_{i,j} - B_{k,j} &= -\frac{1}{2}(F_{i,j} - F_{k,j}) + g_{i,k}(\E),
\end{align*}
where $g_{i, k}(\E) = \frac{1}{2m}(\sum_{t=1}^m E_{i,t} - E_{k,t})$ is a function that depends only on the $\E$ block. 
\end{theorem}
\begin{proof}
First, note that $B_{i,j} = \L(i,:) \tilde{\f}_j$ and $B_{k,j} = \L(k,:) \tilde{\f}_j$. Using these equations, $B_{i,j} - B_{k, j}$ can be expressed as follows: 
\begin{align*}
    B_{i,j} - B_{k, j}    &= (\L(i,:) - \L(k,:)) \tilde{\f}_j \\
    &= m(\e_i - \e_k)^{\top} \tilde{\f}_j \\
    &= -\frac{1}{2}\bg((F_{i,j} - \frac{1}{m}\sum_{t=1}^m E_{i,t}) - (F_{k,j} - \frac{1}{m} \sum_{t=1} E_{k,t}) \bg) \\
    &= -\frac{1}{2}(F_{i,j} - F_{k,j}) + g_{i,k}(\E).
\end{align*} 
\end{proof}

\subsection{Dual Basis Approach for Complete $\E$ and Partially-Observed $\F$}

In this section, we consider the problem where we have complete $\E$ and partially observed $\F$. Despite knowing the $\E$ block completely, we note that this problem makes less stringent assumptions than knowing the positions of the anchors. In particular, using classical MDS on $\E$, we can only recover the configuration of the anchor nodes as opposed to absolute positions. We refer the reader to Appendix B which discusses the case where the positions of the anchors are known. Therein, it is established that the problem has two distinct regimes.  

\begin{figure}[h!]
    \centering
    \includegraphics{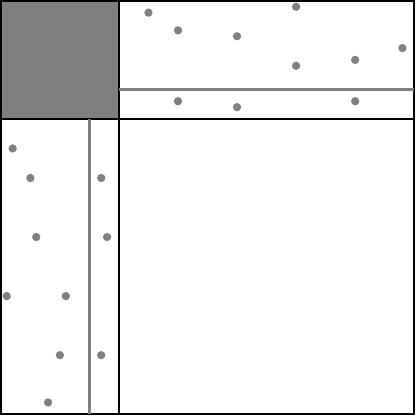}
    \caption{Illustration of the  sampling model with complete anchor-anchor distances and partial observations for anchor-mobile distances. The distances between the mobile nodes are not sampled. In addition, we assume that we know the distance of all mobile nodes from one of the anchors. }
    \label{fig:sampling_illustration_full}
\end{figure}

We now proceed to consider the localization problem for complete $\E$ and partially observed $\F$. The simple relationship between differences of entries in $\B$ and $\F$ in Theorem \ref{b_f_entries_cor} leads us to propose a dual basis framework for this problem. We first consider the case where we have complete anchor-anchor distances but partial anchor-mobile distances. Specifically, the $\E$ block is complete and we have partially-observed entries in $\F$. 
In addition, we possess complete knowledge of all the entries in one specific row of $\F$. Without loss of generality, let us assume this row is row $m$, while all other entries of $\F$ are observed randomly. 
Figure \ref{fig:sampling_illustration_full} illustrates our sampling model. 

Then, motivated by Theorem \ref{b_f_entries_cor},  we can consider our observations to instead be  
\begin{equation}
\label{eq:tildef}
    \tilde{F}_{i,j} = -\frac{1}{2}(F_{i,j} - F_{m,j}) + g_{i,m}(\E),
\end{equation}
for every $i \in [1, m-1]$. Additionally, it is worth noting that $\B$ is characterized by $(m-1)n$ parameters, with a zero column sum constraint. To recover $\B$ from the incomplete observations, we propose the following dual basis approach. First, consider a basis $\w_{i,j} = \e_{i,j} - \e_{m,j}$ that enforces the column sum constraint. Taking an inner product with $\B$ would thus give us 
\begin{align*}
    \langle \B, \tilde{\w}_{i,j} \rangle = B_{i,j} - B_{m,j} 
    = \tilde{F}_{i,j},
\end{align*}
which are exactly the observations.

\begin{theorem}
\label{
dual_basis_partial_f}
Let the set of matrices $\{\tilde{\w}_{i,j} = \e_{i,j} - \e_{m,j}$: $1 \leq i \leq (m - 1), 1 \leq j \leq n\}$ be a basis of the $m \times n$ matrices with zero column sum. 
Then the dual basis vectors are $\{\tilde{\v}_{i,j} = \e_{i,j} - \frac{1}{m}\ones_j: 1 \leq i \leq (m - 1), 1 \leq j \leq n\}$ where $\ones_j$ is a matrix of zeros except a vector of ones at the $j$-th column.
\end{theorem}
\begin{proof}
We can see that the sets $\{ \tilde{\w}_{i,j} \}$ and $\{ \tilde{\v}_{i,j} \}$ both consist of $(m-1)n$ column-centered matrices, each of size $m \times n$.
We will show that they are biorthogonal sets.
We have $\tilde{\w}_{i,j} = \e_{i,j} - \e_{m,j}$ and $\tilde{\v}_{i,j} = \e_{i,j} - \frac{1}{m}\ones_j$, for $1 \leq i < m$, $1 \leq j \leq n$.  If $j \neq l$, then we see immediately that $\langle \tilde{\w}_{i,j}, \tilde{\v}_{k, l} \rangle = 0$. Otherwise, 
\begin{align*}
    \langle \tilde{\w}_{i,j}, \tilde{\v}_{k,j} \rangle &= \langle \e_{i,j} - \e_{m,j}, \e_{k,j} - \frac{1}{m}\ones_j \rangle
    = \delta_i^k.
\end{align*}
Hence, $\langle \tilde{\w}_{\alphab}, \tilde{\v}_{\betab} \rangle = \delta_{\alphab}^{\betab}$, which establishes biorthogonality.
\end{proof}
Here is a quick observation regarding $\H$, the inner product matrix. Given the significant role played by the Laplacian of the complete graph in a preceding section, it is unsurprising that the complete graph resurfaces once more.

\begin{theorem}
\label{h_partial_f}
Consider the set of basis matrices $\{\tilde{\w}_{i,j} = \e_{i,j} - \e_{m,j}$ $(1 \leq i \leq (m - 1), 1 \leq j \leq n)\}$. Then their inner product matrix $\H$, defined by $\H_{\alphab, \betab} = \langle \tilde{\w}_{\alphab}, \tilde{\w}_{\betab} \rangle$,
has the adjacency structure of $n$ disconnected complete graphs on $m-1$ vertices each. 
\end{theorem}

\begin{proof}
We have that $\H_{\alphab, \betab} = 2$ if $\alphab = \betab$, $\H_{\alphab, \betab} = 1$ if $\alphab$ and $\betab$ index entries in the same column (have the same $j$ value), and 0 otherwise. From this we can see that each index $j$ produces $(m-1)$ indices $\alphab$ that all have edges to each other in $\H$ and no edges to any other index $\betab$.
\end{proof}
\subsection{Dual Basis Approach for Incomplete $\E$ and Partially-Observed $\F$}

In the previous section, we discussed a dual basis approach for a complete $\E$ and a partially observed
$\F$. In this section, we consider our main sampling model, the modified Nystr\"om model, and propose
a dual basis approach to recover the submatrix $[\A\, \B]$ from observed distances in $\E$ and $\F$.
The fact that $\E$ is incomplete means that the term $g_{i,m}(\E)$ in \eqref{eq:tildef} is 
not known. However, from the dual basis approach for EDG discussed in Section \ref{sec:dual_basis_approach}, we can relate
$\E$ and $\A$. In particular, $E_{i,j} = \langle \A, \w_{i,j} \rangle = A_{i,i}+A_{j,j}-A_{i,j}-A_{j,i}$. Then, the measurements can be written as follows:
\begin{align*}
    \langle \B, \tilde{\w}_{i,j} \rangle -  \frac{1}{2m}\sum_{t=1}^{m} (A_{i,i}+A_{t,t}-A_{i,t}-A_{t,i})  = -\frac{1}{2}(F_{i,j} - F_{m,j}) 
    -\frac{1}{2m}\sum_{t=1}^{m} E_{m,t}
\end{align*}

\subsection{Proposed Algorithms}

Based on the dual basis formulation, we propose two algorithms for the localization problem
under modified Nystr\"om sampling. The first algorithm considers the case where the submatrix
$\E$ is complete, and we have partial distance information for $\F$. For this case, we consider
the following convex optimization program to recover $\B$:
\begin{equation}\label{eq:nnm_nystrom_B}
\begin{split}
\minimize_{\Z \in \real^{m\times n}} \quad &||\Z||_{*}\\
 \subjectto\quad & Z_{i,j} - Z_{m,j}  = -\frac{1}{2}(F_{i,j} - F_{m,j}) + g_{i,m}(\E) \quad \forall (i,j)\in \Omega\\
 & \Z^T\bm{1}= \bm{0}.\\
\end{split}
\end{equation}
In the above program, $\Omega$ denotes the set of sampled indices in the $\F$ block of the squared
distance matrix $\D$. Let $\B_*$ denote the optimal solution obtained from the above program.
We can then estimate $\C$ using the standard Nystr\"om method $\C = \B_*^T\A^{\dagger}\B_*$.
The second algorithm considers the case where both the submatrix
$\E$ and $\F$ are incomplete, and we only have partial distance information. For this case, we consider
the following convex optimization program to recover $\A$ and $\B$:
\begin{equation}\label{eq:nnm_nystrom_AB}
\begin{split}
\minimize_{\A \in \real^{m\times m},\B \in \real^{m\times n}} \quad &||[\A\,\B]||_{*}\\
 \subjectto\quad & B_{i,j} - B_{m,j} - \frac{1}{2m}\sum_{t=1}^{m} \langle \A,\w_{i,t}\rangle   = -\frac{1}{2}(F_{i,j} - F_{m,j}) -\frac{1}{2m}\sum_{t=1}^{m} E_{m,t} \,\, \,\forall (i,j)\in \Omega_1\\
  \quad & A_{i,i} + A_{j,j} -A_{i,j}-A_{j,i}  = E_{ij} \,\, \,  \forall (i,j)\in \Omega_2.\\
& \A\succeq \bm{0}\\
& \A\bm{1}=\bm{0}, \B^T  \bm{1}=\bm{0}. 
\end{split}
\end{equation}
In the above program, the constraint $\A\bm{1}=\bm{0}$ follows from the representation of $\A$ in the dual basis framework (see Section III-C) and $\Omega_1$ and $\Omega_2$ denote the set of sampled indices in the $\E$ and $\F$ blocks of the squared
distance matrix $\D$, respectively. We note that the above optimization program can also be compactly represented using a single optimization variable $\H=[\A\,\B]$, by appropriately indexing $\A$ and $\B$ within $\H$. Our numerical implementation will be based on this compact representation. Let $\A_*$
and $\B_*$ denote the optimal solutions obtained from the above program.
We can then estimate $\C$ using the standard Nystr\"om method $\C = \B_*^T\A_*^{\dagger}\B_*$.

\section{Dual Basis Centered Anywhere}

\begin{figure}
    \centering
    \includegraphics[scale=0.5]{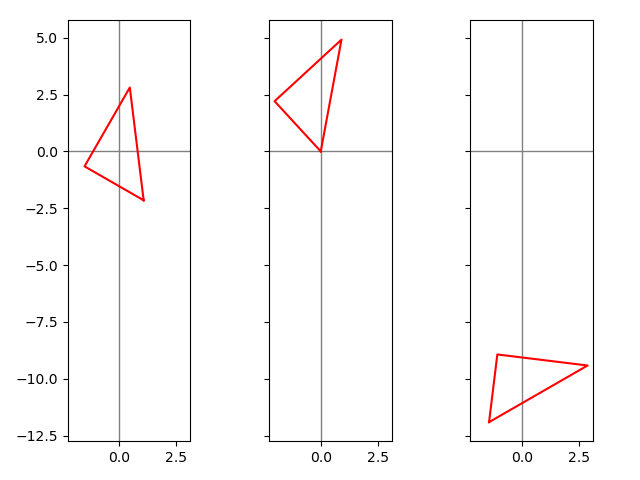}
    \caption{Recovered points $\Pb$ from the same distances $\{3, 4, 5\}$, but with different choices of $\s$: $(1/3, 1/3, 1/3)$, $(1, 0, 0)$, and $(-3, 4, 0)$, respectively. The squared distances of recovered points from the origin are given by $-\frac{1}{2}(\s^{\top}\D\s)\ones + \D\s$ \citep{gower1985properties}.}
    \label{fig:gower-s}
\end{figure}

In \citep{abiy_exact}, the construction of the dual basis framework assumes that the points are centered at $\bm{0}$. In other words, $\P \bm{1} = \bm{0}$. While this simplifies calculations, it is not a mandatory condition for the dual basis approach. As we have observed, there are instances, such as in Nystr\"om setups, where alternative assumptions about the centering of points may be necessary. Fortunately, as shown in \citep{gower1982euclidean,gower1985properties}, for any vector $\s$ that sums to one, we have that 
\begin{align*}
    \Pb^{\top}\Pb = -\frac{1}{2}(\I - \ones \s^{\top})\D(\I - \s \ones^{\top}),
\end{align*}
where $\D$ is the squared Euclidean distance matrix associated to the set of points. 

Under a choice of $\s$, the origin of the points $\Pb$ becomes such that $\Pb\s = \zeros$. 
Here, the vector $\s$ represents a kind of barycentric weighting of each recovered point (as given by a column in $\Pb$). 
We note that the entries of $\s$ are not required to be nonnegative. Figure \ref{fig:gower-s} shows the effect of $\s$ given the set of distances $\{ 3, 4, 5 \}$. We will show that any choice of $\s$ can be used to construct a dual basis setup.

\begin{theorem}
\label{basis_vec}
\label{thm:basis_vec}
Consider a squared Euclidean distance matrix $\D$ and a vector $\s$ such that  $\s^{\top}\ones = 1$. 
Let $\mathbb{S}$ denote the subset of Gram matrices $\X$ that satisfy $\X \s = \zeros$. Let $T$ denote the subspace spanned by $\s$. A basis $\{ \w_{i,j} \}_{i < j}$ of $\mathbb{S}$ satisfying $\langle \w_{i,j}, \X \rangle = D_{i,j}$ and $\X \s = 0$ for all $i < j$ is given by
\begin{align*}
    \w_{i,j} = \a \a^{\top},
\end{align*}
where 
\begin{align*}
    \a &= \mathcal{P}_{T^{\perp}}(\e_i) - \mathcal{P}_{T^{\perp}}(\e_j) 
    = (\e_i - \mathcal{P}_{T}(\e_i)) - (\e_j - \mathcal{P}_{T}(\e_j)) 
    = (\e_i - \frac{(s_i - s_j)}{\|\s\|^2}\s - \e_j).
\end{align*}
\end{theorem}
\begin{proof} It can be readily verified that $\w_{i,j}$ is symmetric and positive semidefinite. By construction,  $\a \in T^{\perp}$. Therefore, any $\X$ expanded in this basis will immediately satisfy $\X \s = \zeros$. 
To show that we recover $D_{i,j}$ via inner products, we first observe that 
\begin{align}
\label{eq:x_wij}
    \langle \X, \w_{i,j} \rangle = \langle -\frac{1}{2}(\I - \ones \s^{\top})\D(\I - \s \ones^{\top}), \w_{i,j} \rangle 
    = -\frac{1}{2}\langle \D, (\I - \s \ones^{\top})\w_{i,j}(\I - \ones \s^{\top}) \rangle,  
\end{align}
by using the trace form of the inner product, followed by the cyclic property of the trace. Consider now the following:
\begin{align*}
    (\I - \s \ones^{\top})(\e_i - \frac{(s_i - s_j)}{\|\s\|^2}\s - \e_j) 
    = (\e_i - \frac{(s_i - s_j)}{\|\s\|^2}\s - \e_j) - (\s - \frac{(s_i - s_j)}{\|\s\|^2}\s - \s) 
    = \e_i - \e_j.
\end{align*}
So therefore the right term in the inner product in \eqref{eq:x_wij} is $(\e_i - \e_j)(\e_i - \e_j)^{\top}$. With that, we do recover the squared distance $D_{i,j}$  as $\langle \X, \w_{i,j} \rangle$. 

Let $\W$ denote the $L \times N^2$ matrix of vectorized basis matrices $\w_{\alphab}$ and, similarly, let $\V$ represent the same for the dual basis matrices $\v_{\alphab}$. Due to biorthogonality (as elaborated in the next Claim), we have $\W \V^{\top} = \I_L$. As $\W$ has a right inverse, it must be full row rank, thereby establishing that the space spanned by $\{ \w_{\alphab}\}$ has dimension $L$ as desired.

\end{proof}
 
As a remark, we observe that, when $s_i = s_j$ for all $i,j$, then we recover the form of the basis of \citep{abiy_exact}. This basis expansion thus is a strict generalization. To complete the dual basis framework, we must also provide the dual basis vectors, and show that our two sets of vectors comprise a biorthogonal system. 

\begin{theorem}
\label{dual_vec}
Given $\w_{i,j}$ under a choice of $\s$ such that $\s^{\top} \ones = 1$, the dual basis vectors are given by
\begin{align*}
   \v_{i,j} = -\frac{1}{2}(\c \d^{\top} + \d \c^{\top})
\end{align*}
where $\c = (\e_i - s_i \ones)$ and $\d = (\e_j - s_j \ones)$.
\end{theorem}
The proof is deferred to the Appendix.

\subsection{Equivalence of Dual Basis Approach with Nystr\"om}

In this section, we show that we can also recover the Nystr\"om method via a dual basis expansion of $\K$. Recall that we have the expansion
\begin{align*}
    \K &= \sum_{\alphab \in \univ} \langle \K, \w_{\alphab} \rangle \v_{\alphab}. 
\end{align*}
By restricting the dual basis vectors to particular blocks, we thus obtain the relationships
\begin{align*}
    \A = \sum_{\alphab \in \univ} \langle \K, \w_{\alphab} \rangle  \v_{\alphab}^{(1,1)} ,\quad 
    \B = \sum_{\alphab \in \univ} \langle \K, \w_{\alphab} \rangle  \v_{\alphab}^{(1,2)}. 
\end{align*}
In what follows, for ease of notation, we define $\bar{\e}_i$ and $\bar{e}_j$ as follows:
\begin{align*}
\bar{\e}_i \equiv \e_i - \frac{1}{m}\ones\\
\bar{\e}_j \equiv \e_j - \frac{1}{m}\ones.
\end{align*}
Under Claim 2.2, for a Nystr\"om choice of $\s$, we have that 
\begin{align*}
    \v_{i,j} 
    &= -\frac{1}{2}\begin{cases}
        \bar{\e_i}\bar{\e_j}^T+\bar{\e_j}\bar{\e_i}^T
        & \text{if $i < j \leq m$} \\
        \bar{\e_i}\bar{\e_j}^{\top} + \e_j \bar{\e_i}^{\top} & \text{if $i \leq m < j$} \\
        \e_i\e_j^{\top} + \e_j \e_i^{\top} & \text{if $m < i < j$}
    \end{cases}
\end{align*}
We see then that $\v_{i,j}^{(1,1)}$ is the zero matrix if $m < j$, and $\v_{i,j}^{(1,2)}$ is the zero matrix if $m < i < j$. Therefore we have 
\begin{align*}
    \A &= \sum_{i < j \leq m} \langle \K, \w_{\alphab} \rangle  \v_{\alphab}^{(1,1)} \\
    &= \sum_{i < j \leq m} E_{i,j} \v_{i,j}^{(1,1)} \\
    &= -\frac{1}{2}\sum_{i < j \leq m} E_{i,j} \bg[\bar{\e_i}\bar{\e_j}^{\top}+\bar{\e_j}\bar{\e_i}^{\top} \bg]\\
     &= -\frac{1}{2} \bg(\sum_{i=1}^m \sum_{j=1}^m E_{i,j} \bar{\e_i}\bar{\e_j}^{\top} \bg), 
\end{align*}
where the last equality follows since $E_{i,i} = 0$. Looking at a particular index $(s,t)$, we have that
\begin{align*}
    A_{s,t}  &= -\frac{1}{2}\bg(\sum_{i=1}^m \sum_{j=1}^m E_{i,j}(\delta_i^s - \frac{1}{m})(\delta_j^t - \frac{1}{m})^{\top} \bg) \\
    &= -\frac{1}{2}\bg(\sum_{i=1}^m \sum_{j=1}^m E_{i,j}(\delta_i^s \delta_j^t - \delta_i^s\frac{1}{m} - \delta_j^t \frac{1}{m} + \frac{1}{m^2}) \bg) \\
    &= -\frac{1}{2}\bg(E_{s,t} - \frac{1}{m}\sum_{j=1}^m E_{s,j} - \frac{1}{m}\sum_{i=1}^m E_{i,t} + \frac{1}{m^2} \sum_{i=1}^m \sum_{j=1}^m E_{i,j}\bg).
\end{align*}
The $\B$ block is a little trickier since we have to only consider the terms of $\v_{i,j}$ that appear in the $(1,2)$ block.
\begin{align*}
    \B &= \sum_{i < j \leq m} \langle \K, \w_{\alphab} \rangle  \v_{\alphab}^{(1,2)} + \sum_{i \leq m < j} \langle \K, \w_{\alphab} \rangle  \v_{\alphab}^{(1,2)} \\
    &= \sum_{i < j \leq m} E_{i,j} \v_{i,j}^{(1,2)} + \sum_{i \leq m < j} F_{i,j} \v_{i,j}^{(1,2)} \\
    &= -\frac{1}{2}\bg(\sum_{i < j \leq m} E_{i,j}
    \bg[(\e_i-\frac{1}{m}\ones)( - \frac{1}{m}\ones)^{\top} + ( \e_j-\frac{1}{m}\ones)(-\frac{1}{m}\ones)^{\top}\bg]    
    + \sum_{i \leq m < j} F_{i,j} 
    \bg[(\e_i - \frac{1}{m}\ones)\e_j^{\top}\bg]\bg) \\
    &= -\frac{1}{2}\bg(\sum_{i=1}^m \sum_{j=1}^m -\frac{1}{m}E_{i,j}
    \bar{\e}_i(\ones)^{\top}
    + \sum_{i \leq m < j} F_{i,j} 
    \bg[\bar{\e}_i\e_j^{\top}\bg]\bg).
\end{align*}
A particular choice of index $s,t$ yields 
\begin{align*}
    B_{s,t} 
    &= -\frac{1}{2}\bg( 
    \sum_{i=1}^m \sum_{j=1}^m E_{i,j} (-\delta_i^s\frac{1}{m} + \frac{1}{m^2})
     + \sum_{i \leq m < j} F_{i,j} (\delta_i^s\delta_j^t - \frac{1}{m}\delta_j^t) \bg) \\
     &= -\frac{1}{2}\bg( 
     F_{s,t} - \frac{1}{m}\sum_{i=1}^m F_{i,t} - \frac{1}{m} \sum_{j=1}^m E_{s,j} + \frac{1}{m^2}\sum_{i=1}^m \sum_{j=1}^m E_{i,j}.
     \bg)
\end{align*}
So we recover $\A$ and $\B$ only in terms of $\E$ and $\F$, with exactly the same relationship as in the Nystr\"om method.

\section{Results}
In this section, we test Algorithm 2 on both synthetic and real datasets. To solve Algorithm 2, we use CVX \citep{cvx,gb08}. The specific solver we used is Mosek 10.2.1 \citep{mosek}. For both experiments, our sampling scheme for the distance matrices we generate is as follows. Initially, the last row of $\E$ and $\F$ are fully sampled, corresponding to the central node.
The remaining distance entries in $\E$ block are sampled according to the Bernoulli sampling model, with each entry having a probability of $\gamma$ being selected, and where $\gamma \in [0,1]$. Note that the diagonal entries of $\E$, which are all zero, are assumed to be known as $\E$ is a valid distance matrix. For each column of
$\F$, besides its $m$-th entry, we sample $\alpha-1$ entries uniformly at random without replacement. We adopt this strategy to study the effect of samples in $\E$ vs. samples in $\F$, and because we empirically observed that recovery for low rank depends much more on the entries in $\F$.

\subsection{Experiments on Synthetic Data}
For the synthetic experiments, we set $n = 500, m = 50, r = 2$, and generate a matrix $\P$ of $(m + n)$ random points in $\real^2$, where each coordinate has an i.i.d standard normal distribution. We then subtract off the mean of the first $m$ points (to reproduce the centering in \citep{platt2004nystrom}, which our dual basis assumes holds). From these points $\P$, we produce a squared distance matrix $\D$. We complete the $[\A\,\B ]$ block of the Gram matrix $\P^{\top} \P$ using a single convex optimization program (Algorithm 2) that merges the constraints from both the dual basis and semidefinite constraint on the $\E \to \A$ block as well as the constraints arising from the dual basis we introduced in this paper for the $\F \to \B$ block.  We then use the standard Nystr\"om method to produce an estimate for the full Gram matrix, and do a truncated eigendecomposition and Procrustes analysis to produce an RMSE that compares our estimate against the original points $\P$.

For each choice of $\gamma$ and $\alpha$, we calculate the mean and standard deviation of the RMSE across 25 trials. We attempted to compare with a standard SDP to solve for $\X$ from $\D$, but the semidefinite constraint in such a problem involves a much larger matrix $(m + n) \times (m + n)$ vs $m \times m$, so the memory and time requirements quickly became intractable, requiring several hundreds of GB of memory in CVX to run a single trial for sizes of $n$ over 200.

\begin{table}[h]
\renewcommand{\arraystretch}{1.25}
\caption{Synthetic experiment results.}
\label{tab:my_table}
\centering
\begin{tabular}{|c|c|c|c|}
\hline
$\gamma$ (sampling rate in $\E$) & $\alpha$ (samples in each $\F$ column) & Average RMSE & Standard deviation \\ \hline
0.5 & 10 & $1 \times 10^{-11}$ & $1 \times 10^{-11}$ \\ \hline
0.3 & 10 & $2.2 \times 10^{-3}$ & $9.1 \times 10^{-3} $ \\ \hline
0.1 & 10 & $1.9 \times 10^{-3}$ & $5.7 \times 10^{-3} $ \\ \hline
0.0 & 10 & $6.04 \times 10^{-4}$ & $2.8 \times 10^{-3}$ \\ \hline
0.5 & 8 & $9.3 \times 10^{-3} $ & $2.04 \times 10^{-2}$ \\ \hline
0.5 & 5 & $2.3 \times 10^{-1}$ & $7.11 \times 10^{-2}$ \\ \hline
\end{tabular}

\end{table}

\subsection{Experiments on Real Data}
\begin{figure}[h!]
    \centering
    \includegraphics[scale=0.45]{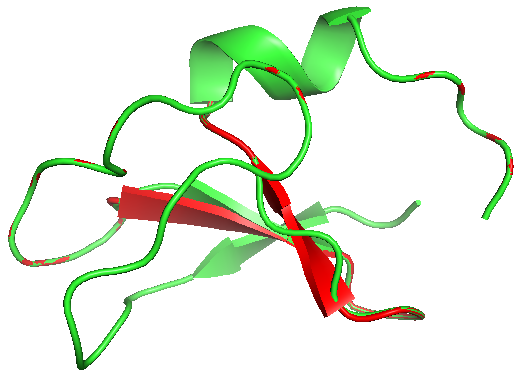}
    \caption{Target structure 1PTQ (in green) and numerically estimated structure (in red). A realization from the experiment with $\gamma = 0.2$ and $\alpha=6$ (RMSE = 0.65). }
    \label{fig:protein_comparisons}
\end{figure}

For our real data experiments, we evaluate our main algorithm, Algorithm 2, on protein data identified as 1PTQ. The protein data is obtained from the Protein Data Bank \citep{berman2000protein}. 1PTQ has 402 atoms and we select $m=20$ anchors uniformly spaced within the
range $[1,402]$. It is worth noting that, in practice, more refined methods for selecting anchors can be employed, using domain knowledge. We then run Algorithm 2 and assess the quality of the resulting numerical solution using the root mean square error (RMSE) between the numerically estimated protein structure and the ground truth. Since the sampling in the $\F$ block is random, we report the average RMSE from a total of 25 trials. Table \ref{tab:protein_results} summarizes the results of our numerical experiments. We note that, if there are sufficient distance measurements in the $\F$ block, the structure can be estimated with very high accuracy. In our numerical experiments, similar to the discussion in the synthetic experiments, the number of samples in the $\F$ block influences the final RMSE significantly compared to the sampling rate in $\E$. Figure \ref{fig:protein_comparisons} shows a visualization that compares the true structure with estimated structure. The visualization was generated using the PyMOL Molecular Graphics System \citep{PyMOL}. 

It is important to recognize that a realistic sampling scheme for distances in proteins is based on the proximity of atoms rather than uniform sampling. The numerical experiment we discussed aims to apply the proposed algorithm to real data. In practice, implementing this approach may require an initial estimate for the target structure and a domain knowledge of which atoms serve as reliable anchors.

\begin{table}[h!]
\renewcommand{\arraystretch}{1.25}
\caption{Average RMSE and standard deviation for different total numbers of sampled entries per column of $\F$. $m$ is set to $20$ and $\gamma = 0.2$.}
\label{tab:protein_results}
\centering
\begin{tabular}{|c|c|c|}
\hline
$\alpha$ (samples in each $\F$ column) & Average RMSE &  Standard deviation \\ \hline
7 & 1.144 &  $7.910\times 10^{-1}$ \\ \hline
8 & $2.002\times 10^{-1}$ &  $3.655\times 10^{-1}$ \\ \hline
9 & $1.51\times 10^{-2}$ &  $4.75\times 10^{-2}$ \\ \hline
10 & $3.944\times 10^{-10}$ & $8.653\times 10^{-10}$ \\ \hline
\end{tabular}

\end{table}

\begin{figure}[h!]
    \centering
    \includegraphics[scale=0.53]{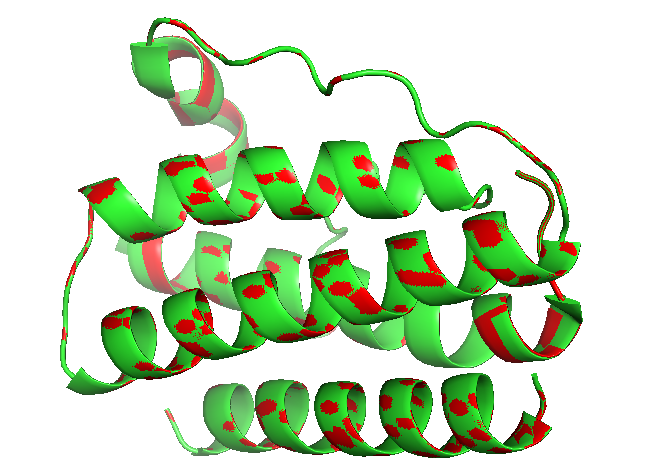}
    \caption{Target structure 1AX8 (in green) and numerically estimated structure (in red). A realization from the experiment with $\gamma = 0.2$ and $\alpha=9$ (RMSE = 0.17). }
    \label{fig:protein_comparisons_2}
\end{figure}

Next, we apply our method to a larger protein identified as 1AX8 from the Protein Data Bank \citep{berman2000protein}. 1AX8 has 1003 atoms, ignoring the hetero atoms, and we select $m=30$ anchors uniformly spaced within the
range $[1,1003]$. Figure \ref{fig:protein_comparisons_2} shows one realization of our numerical experiment, which shows that the underlying structure can be estimated exactly.                                 

\section{Conclusion}

In this paper, we study the problem of determining the configuration of $n$ points using only partial distance information relative to  $m$ points, along with incomplete distance information between the $m$ points. For this setup, we first establish a connection between the distance matrix and the Gram matrix that only utilizes the observed blocks of the distance matrix. We then demonstrate that the problem of recovering the positions can be framed as a low-rank recovery of a submatrix of the Gram matrix. We test the proposed method on synthetic and real data, showing that it can accurately estimate the configuration of the points given reliable anchors and sufficient distance samples. Future research will focus on quantifying ``reliable anchors" and providing theoretical guarantees for the proposed optimization.

\section*{Acknowledgements}

Abiy Tasissa acknowledges partial support from NSF through grant DMS-2208392.  Abiy Tasissa would like to thank Professor Alex Gittens and Prof. Farhad Pourkamali-Anaraki for discussions on the Nystr\"om method.

\bibliographystyle{IEEEtranN}
\bibliography{IEEEabrv,refs}

\newpage
\appendix

\section{Proof of Theorem \ref{dual_vec} (Establishing Biorthogonality)}
\begin{proof}
We need to show that $\langle \w_{i,j}, \v_{i,j} \rangle = 1$ and $\langle \w_{i,j}, \v_{s,t} \rangle = 0$ for $(i,j) \neq (s,t)$. We have three cases to check: same indices, no indices shared, one index shared. 
\snl 

\noindent \underline{Case 1: Same indices}. We are interested in the inner product 
\begin{align*}
\langle \w_{i,j}, \v_{i,j} \rangle 
& = -\frac{1}{2} \langle \a \a^{\top}, (\c \d^{\top} + \d \c^{\top}) \rangle \\
&= 
-\frac{1}{2}\text{Trace}(\a \a^{\top} (\c \d^{\top} + \d \c^{\top})) \\
&= 
-\frac{1}{2}\bg(\text{Trace}(\a \a^{\top}\c \d^{\top}) + \text{Trace}(\a \a^{\top}\d \c^{\top})\bg),
\end{align*}
where we have that $\a = (\e_i - \frac{(s_i-s_j)}{\|s\|^2}\s - \e_j)$, $\c = (\e_i - s_i \ones), \d = (\e_j - s_j \ones)$. 
First we look at
\begin{align*}
   \a^{\top} \c  & =
    (\e_i - \frac{(s_i - s_j)}{\|\s\|^2}\s - \e_j)^{\top}(\e_i - s_i \ones ) \\
   &= \e_i^{\top}\e_i - \frac{(s_i - s_j)}{\|\s\|^2}s_i - \e_j^{\top}\e_i - s_i + \frac{(s_i - s_j)}{\|\s\|^2}s_i + s_i \\
   &= 1,
\end{align*}
and
\begin{align*}
    \a^{\top} \d & = (\e_i - \frac{(s_i - s_j)}{\|\s\|^2}\s - \e_j)^{\top}(\e_j - s_j \ones ) \\
   &= \e_i^{\top}\e_j - \frac{(s_i - s_j)}{\|\s\|^2}s_j - \e_j^{\top}\e_j - s_j + \frac{(s_i - s_j)}{\|\s\|^2}s_j + s_j \\
   &= -1.
\end{align*}
Therefore we have
\begin{align*}
    \langle \w_{i,j}, \v_{i,j} \rangle = -\frac{1}{2}\bg(\text{Trace}(\a \d^{\top}) - \text{Trace}(\a \c^{\top}) \bg).
\end{align*}
We have that
\begin{align*}
    \text{Trace}(\a \d^{\top}) &= \sum_{k=1}^n (\e_i - \frac{(s_i - s_j)}{\|\s\|^2}\s - \e_j)_k (e_j - s_j \ones)_k \\
    &= \sum_{k=1}^n (\delta_i^k - \frac{(s_i - s_j)}{\|\s\|^2}s_k - \delta_j^k)(\delta_j^k - s_j) \\
    &= \sum_{k=1}^n \bg( \delta_i^k \delta_j^k - \delta_i^k s_j -  \frac{(s_i - s_j)}{\|\s\|^2}s_k \delta_j^k + \frac{(s_i - s_j)}{\|\s\|^2}s_k s_j \\
    &- \delta_j^k \delta_j^k + \delta_j^k s_j \bg) \\
    &= -s_j - \frac{(s_i - s_j)}{\|\s\|^2}s_j + \frac{(s_i - s_j)}{\|\s\|^2}s_j -1 + s_j \\
    &= -1,
\end{align*}
and 
\begin{align*}
     \text{Trace}(\a \c^{\top}) &= \sum_{k=1}^n (\e_i - \frac{(s_i - s_j)}{\|\s\|^2}\s - \e_j)_k (e_i - s_i \ones)_k \\
    &= \sum_{k=1}^n (\delta_i^k - \frac{(s_i - s_j)}{\|\s\|^2}s_k - \delta_j^k)(\delta_i^k - s_i) \\
    &= \sum_{k=1}^n \bg( \delta_i^k \delta_i^k - \delta_i^k s_i -  \frac{(s_i - s_j)}{\|\s\|^2}s_k \delta_i^k + \frac{(s_i - s_j)}{\|\s\|^2}s_k s_i \\
    &- \delta_j^k \delta_i^k + \delta_j^k s_i \bg) \\
    &= 1 -s_i - \frac{(s_i - s_j)}{\|\s\|^2}s_i + \frac{(s_i - s_j)}{\|\s\|^2}s_i  + s_i \\
    &= 1,   
\end{align*}
so we conclude that
\begin{align*}
    \langle \w_{i,j}, \v_{i,j} \rangle &= 
    -\frac{1}{2}\bg(\text{Trace}(\a \d^{\top}) - \text{Trace}(\a \c^{\top}) \bg)= -\frac{1}{2}\bg(-1 - 1 \bg)     = 1
\end{align*}
which is as desired.
\snl 

\noindent \underline{Case 2: no index shared}. Suppose that $(i,j)$ and $(p,q)$ have no index in common. For this case, we have that $\a = (\e_i - \frac{(s_i-s_j)}{\|s\|^2}\s - \e_j)$, $\c = (\e_p - s_p \ones), \d = (\e_q - s_q \ones)$. We see that
\begin{align*}
    \a^{\top}\c &= (\e_i - \frac{(s_i - s_j)}{\|\s\|^2}\s - \e_j)^{\top}(\e_p - s_p \ones) = 0
\end{align*}
since $\s^{\top} \ones = 1$, and likewise $\a^{\top} \d$ will also be zero. So the inner product $\langle \w_{i,j}, \v_{p,q} \rangle$ is zero, as desired.

\snl 

\noindent \underline{Case 3: one index in common}. Since the matrices are symmetric, suppose without loss of generality that the first index is equal: we consider $\langle \w_{i,j}, \v_{i,q} \rangle$, $j \neq q$. 
For this case, we have that $\a = (\e_i - \frac{(s_i-s_j)}{\|s\|^2}\s - \e_j)$, $\c = (\e_i - s_i \ones), \d = (\e_q - s_q \ones)$.
We see that 
\begin{align*}
 \langle \w_{i,j}, \v_{i,q} \rangle &= -\frac{1}{2}\bg(\text{Trace}(\a \a^{\top}\c \d^{\top}) + \text{Trace}(\a \a^{\top}\d \c^{\top})\bg) \\
 &= -\frac{1}{2}\text{Trace}(\a \d^{\top}) 
\end{align*}
since $j \neq q \implies \a^{\top}\d = (\e_i - \frac{(s_i-s_j)}{\|s\|^2}\s - \e_j)^{\top}(e_q-s_q \ones) = 0$, and $\a^{\top} \c = 1$, as the analysis in Case 1 shows.
We examine the trace:
\begin{align*}
    \text{Trace}(\a \d^{\top}) 
    &= \sum_{k=1}^n (\e_i - \frac{(s_i - s_j)}{\|\s\|^2}\s - \e_j)_k (e_q - s_q \ones)_k \\
    &= \sum_{k=1}^n (\delta_i^k - \frac{(s_i - s_j)}{\|\s\|^2}s_k - \delta_j^k)(\delta_q^k - s_q) \\
    &= \sum_{k=1}^n \bg( \delta_i^k \delta_q^k - \delta_i^k s_q -  \frac{(s_i - s_j)}{\|\s\|^2}s_k \delta_q^k + \frac{(s_i - s_j)}{\|\s\|^2}s_k s_q \\
    &-\delta_j^k \delta_q^k + \delta_j^k s_q \bg) \\
    &= -s_q - \frac{(s_i - s_j)}{\|\s\|^2}s_q + \frac{(s_i - s_j)}{\|\s\|^2}s_q + s_q \\
    &= 0,
\end{align*}
which shows us that the inner product is zero.

\snl
This completes all the cases, and proves that we indeed have a biorthogonal system. 
\end{proof}

\section{Localization Problem Formulation Given Exact Anchor Positions}

In this section, we study the localization problem where the exact position of the anchor nodes are known, and partial information about the distances between the anchor nodes and mobile nodes is available. We highlight two regimes of the problem:
\begin{itemize}[leftmargin=*]
\item Determined Regime: If we have distances from $r+1$ affinely independent anchors to a mobile node, the mobile node's position can be localized exactly. This is similar to the trilateration method used in GPS, which determines an object's position based on distances from known locations. For instance, in 2D trilateration, distances to 3 non-degenerate anchors uniquely determine the target point's position \citep{fang1986trilateration,thomas2005revisiting,wang2015linear}.

\item Underdetermined Regime: When we have distances from fewer than $r+1$ anchors, the problem has infinitely many solutions.
\end{itemize}
Let $\Omega \subset
\{(i,j)|1 \leq i \leq (m - 1), 1 \leq j \leq n\}$ denote the set of sampled indices in the $\F$ block of the squared distance matrix. Using our dual basis formulation, we have shown that  $\langle \B,\tilde{\w}_{i,j}\rangle= \tilde{F}_{i,j}$ where $\tilde{F}_{i,j}$ is given in \eqref{eq:tildef}. The recovery problem for $\B$ can be formulated as the following feasibility problem:
\begin{equation}
\label{eq:feasibility_b}
\textrm{ Find } \B\in \real^{m\times n} \,\,\textrm{ such that }\,\,
 \langle \B, \tilde{\w}_{i,j} \rangle = B_{i,j} - B_{m,j}  = -\frac{1}{2}(F_{i,j} - F_{m,j}) + g_{i,m}(\E) \quad \forall (i,j) \in \Omega,
\end{equation}
where $g_{i, m}(\E) = \frac{1}{2m}(\sum_{t=1}^m E_{i,t} - E_{m,t})$ is the function that depends only on the $\E$ block. Since $\B=\X^T\Y$, the above problem can also be stated equivalently as follows:   
\begin{equation}
\label{eq:feasbility_y}
    \textrm{ Find } \Y\in \real^{r\times n} \,\,\textrm{ such that }\,\,
 \langle \Y, \X\bm{z}_{i,j} \rangle = -\frac{1}{2}(F_{i,j} - F_{m,j}) + g_{i,m}(\E) \quad \forall (i,j) \in \Omega.
\end{equation}
In fact, \eqref{eq:feasbility_y} can be written as $\A_{\Omega}(\Y)=\b_{\Omega}$ where $\A_{\Omega}$ is a linear operator. This is a linear system problem. 
We establish conditions under which the linear system in \eqref{eq:nnm_nystrom_y} admits a unique solution.

\begin{theorem}
\label{unique_solution}
If we observe $r+1$ distance measurements in each column of $F$ and if any $r+1$ anchors are affinely independent, then \eqref{eq:feasbility_y} has a unique solution, and the mobile nodes can be localized exactly. 
\end{theorem}

\begin{proof}
Given a mobile node $\y_j$, assume that we have $r+1$ distances to the anchors denoted by $\x_{i_1},\x_{i_2},...,\x_{i_m}$. Note the last anchor is $\x_{i_m}=\x_m$ corresponding to the central node from which we know the distance to any mobile node. With that, \eqref{eq:feasbility_y} is equivalent to solving the linear system
\[ 
\begin{pmatrix}
\x_{i_1}-\x_{i_m}\\
\x_{i_2}-\x_{i_m}\\
\vdots\\
\x_{i_r}-\x_{i_m}\\
    \end{pmatrix}
    \begin{pmatrix}
    (\y_j)_1\\
        (\y_j)_2\\
        \vdots\\
            (\y_j)_r\\
            \end{pmatrix}=
    \begin{pmatrix}
\tilde{F}_{i_1,j}\\
\tilde{F}_{i_2,j}\\
\vdots\\
\tilde{F}_{i_r,j}\\
\end{pmatrix}.
    \]
    The system has a unique solution if and only if the the matrix in the above linear system is invertible. We note that the $r$ rows of the matrix are linearly independent if and only if the $r+1$ anchors are affinely independent. Since this is true for any mobile node and the $r+1$ anchor nodes from which we have distance information to, the conclusion of the theorem follows.
\end{proof}
\noindent \textbf{Remark}: As remarked earlier, the above theorem is equivalent to trilateration, which ensures recovery of a mobile node given $r+1$ non-degenerate anchors \citep{fang1986trilateration,thomas2005revisiting,wang2015linear}. 

In the underdetermined regime, we have fewer than $r+1$ entries in each column of $F$, i.e., we know the distance of a mobile node
from fewer than $r+1$ anchors. Theorem \ref{unique_solution} does not apply in this setup, and indeed \eqref{eq:feasbility_y} admits infinitely many solutions. To obtain a unique solution, additional assumptions on the underlying points are necessary. In this paper, we consider points embedded in high dimensions ($\real^{r}$) that are intrinsically low-dimensional ($k$-dimensional), with $k\ll r$. Additionally, we assume that entries of each column of $\F$ are sampled uniformly at random with replacement. We note that, under the low-dimensional assumption of the underlying points, $\B$ is a low-rank matrix.  We propose the following optimization program to estimate $\B$:
\begin{equation}\label{eq:nnm_nystrom}
\begin{split}
\minimize_{\Z \in \real^{m\times n}} \quad &
\lVert\Z\rVert_{F}^2\\
  \subjectto\quad & \langle \Z,\tilde{\w}_{i,j}\rangle = \langle \B,\tilde{\w}_{i,j}\rangle\quad \forall (i,j)\in \Omega.
\end{split}
\end{equation}
Alternatively, using the fact that $\B=\X^T\Y$, we can directly estimate the positions of the mobile nodes by using the following optimization program:
\begin{equation}\label{eq:nnm_nystrom_y}
\begin{split}
\minimize_{\Y \in \real^{r\times n}} \quad &\lVert\Y\rVert_{F}^2\\
 \subjectto\quad & \langle \Y,\X\tilde{\w}_{i,j}\rangle = \langle \B,\tilde{\w}_{i,j}\rangle\quad \forall (i,j)\in \Omega.
\end{split}
\end{equation}
We note that \eqref{eq:nnm_nystrom_y} is a least squares problem, and the solution can be computed in closed form. This solution is known to be provably low rank \citep{zhang2014flrr}.
\end{document}